\documentclass[12pt]{article}

\usepackage{epsf,epsfig}

\newcommand{\e}{\varepsilon}
\newcommand{\p}{\bot}
\newcommand{\pa}{\scriptscriptstyle \|}
\newcommand{\dd}{\partial}
\newcommand{\de}{\delta}
\newcommand{\m}{\mu}
\newcommand{\n}{\nu}
\newcommand{\om}{\omega}
\newcommand{\La}{\Lambda}
\newcommand{\la}{\lambda}
\newcommand{\ls}{\left(}
\newcommand{\rs}{\right)}
\newcommand{\str}[1]{\mathrel{\mathop{\longrightarrow}\limits_{#1}}}
\newcommand{\al}{\alpha}
\newcommand{\be}{\beta}
\newcommand{\tr}{{\rm tr}\!}
\newcommand{\ff}{\varphi}
\newcommand{\ga}{\gamma}
\newcommand{\De}{\Delta}
\newcommand{\ddl}{\acute\partial}
\newcommand{\ddr}{\grave\partial}

\newcommand{\disn}[2]{$$\displaylines{\refstepcounter{equation}%
            \label{#1}\hskip 1em minus 1em #2\hfilneg}$$}
\newcommand{\nom}{\hfil\hskip 1em minus 1em (\theequation)}
\newcommand{\no}{\hfil \hskip 1em minus 1em\phantom{(\theequation)}%
            \hfilneg\cr\hfilneg\hskip 1em minus 1em\hfil}
\newcommand{\ns}{\hfill\cr\hfill}

\begin{document}

\title{
Feynman perturbation theory\\
for gauge theory on transverse lattice\\
}
\author{M.S.~Karnevskiy\thanks{E-mail: karnevsky@gmail.com},
S.A.~Paston\thanks{E-mail: paston@pobox.spbu.ru}\\
{\it Saint Petersburg State University, Russia}
}
\date{\vskip 15mm}
\maketitle

\begin{abstract}
Feynman perturbation theory for nonabelian gauge theory in light-like gauge is investigated.
A lattice along two space-like directions is used as a gauge invariant ultraviolet regularization.
For preservation of the polinomiality of action we use as independent variables arbitrary (non-unitary)
matrices related to the link of the lattice. The action of the theory is selected in such a way to preserve
as much as possible the rotational invariance, which remains after introduction of the lattice,
as well as to make superfluous degrees of freedom vanish in the limit of removing the regularization.
Feynman perturbation theory is constructed and diagrams which does not contain ultraviolet divergences
are analyzed. The scheme of renormalization of this theory is discussed.
\end{abstract}

\newpage
\section{Introduction}
The problem of determining of the mass spectrum of bound states for QCD is
not completely solved yet. The fact that the coupling constant is not small at low
energies motivates the development of nonperturbative calculation methods.
One of such methods is the canonical quantization theory
in the Light Front (LF) coordinates $x^\pm=(x^0\pm x^3)/\sqrt{2}$, $x^1$, $x^2$, where $x^+$
is time\cite{dir}. A canonical LF Hamiltonian is constructed as a result of the quantization. Afterwards, the calculation of its spectrum has to be performed. The advantage of this approach is a simple form of the vacuum
state: the physical vacuum coincides with the mathematical vacuum.
This fact simplifies the calculation of the mass spectrum of bound states (see, for example,  Ref.~\cite{yf05}), however it is necessary to note, that it is strictly correct only when an appropriate regularization is introduced.
The method of quantization in LF
coordinates was described in detail in the review Ref.~\cite{brodsk}.
Also we can note  the recent works\cite{brodskme,brodsk2} which use this method.

Unfortunately, the quantization in LF coordinates may lead to the theory
which is not equivalent to the original theory in Lorentz coordinates\cite{bur,tmf97}.
This is due to appearance of the singularity of the theory at light-like momentum $p_-$, equal zero.
Regularization of this singularity leads to broken Lorentz invariance. However, it
is possible to recover equivalence of two approaches, at least in all orders of the
coupling constant perturbation theory, correcting "naive" LF Hamiltonian
by addition of certain counterterms. This, "corrected", LF Hamiltonian
can be used for nonperturbative calculations. The
general method of finding a "corrected" LF Hamiltonian by analyzing of the Feynman
perturbation theory in all orders is given in Ref.~\cite{tmf97}. An example of successful
application of this method together with numerical calculation of
the spectrum of Hamiltonian in the "Discretized Light-Cone
Quantization" (DLCQ, see Ref.~\cite{brodsk})
method is the investigation of a 2D model,
the massive Schwinger model\cite{tmf02,yf05}. The mass spectrum calculated nonperturbatively with the "corrected" LF
Hamiltonian, was in good agreement with the results of lattice calculations
using usual coordinates for a wide range of fermion mass. At the same time, the
use of "naive" LF Hamiltonian gives correct results only for small masses of a
fermion\cite{yf05}.

Construction of the "corrected" LF Hamiltonian for the 4D nonabelian gauge theory, in particular for QCD, is a very difficult problem. Such a QCD Hamiltonian is constructed in the article Ref.~\cite{tmf99}, but the gauge invariance in this construction is broken. As a result, renormalization leads to a large
number of counterterms with unknown coefficients. It should be noted
that unlike 2D models (where the space of the states is found to be finite-dimensional in the framework of DLCQ method, see e.~g. Ref.~\cite{yf05}), for 4D theory, space of states is not finite-dimensional, because there are additional directions $x^1,x^2$. Therefore, it is difficult to calculate the mass spectrum of LF Hamiltonian.

These two problems, the violation of gauge invariance and the infinite dimension
of the space of states, can be solved at once if the transverse lattice $x^1,x^2$ is introduced.
The usual way of the introduction of the lattice is when the components of the gauge field in the discrete directions are replaced by unitary matrices related to the links of the lattice. Unitarity of these matrices makes action of the theory non-polynomial relatively to
independent variables. This makes analysis of Feynman perturbation theory extremely difficult. Therefore, it seems more promising to consider these matrices as arbitrary complex (see review Ref.~\cite{pirn} for applying this idea to 4D lattice) and use a transverse lattice method as it is proposed in Ref.~\cite{barpir1}. In this approach, action is polynomial and analysis of the perturbation theory is more simple, however, non-physical degrees of freedom appear in the theory.

In order to apply the method of "corrected" LF Hamiltonian to the theory on the transverse lattice, Feynman perturbation theory on the transverse lattice should be formulated. Furthermore, an action should be selected in such a way that the Green functions of the theory in Lorentz coordinates coincide with the Green functions of the normal QCD in the zero limit of lattice spacing $a$ and the non-physical degrees of freedom are switched off from the theory. It is necessary also to perform the procedure of renormalization of the theory with the lattice regularization being removed.
After that a "corrected" LF Hamiltonian can be constructed by the method from Ref.~\cite{tmf97}.
This Hamiltonian would correspond to the normal QCD at least in framework of perturbation theory (in all
orders) by the coupling constant. This LF Hamiltonian can be used for non-perturbative calculation of mass spectrum of QCD. Its main advantage, in comparison to the Hamiltonian which has been constructed in article Ref.~\cite{tmf99}, is a finite-dimensional space of states in the
DLCQ method. In addition, the use of a lattice can give hopes that the number of unknown factors would be smaller, because instead of violating gauge invariance (see a more accurate discussion in Sect.~4), there is only a partial violation of Lorenz invariance.

The transverse lattice method which includes the construction of the LF Hamiltonian and introduction
of the transverse lattice is very useful for description of QCD (see review Ref.~\cite{burdal} and references therein). However, in this approach, the method of arbitrary complex matrices is usually understood as a transition to new "colour-dielectric" variables which correspond to important degrees of freedom on the coarse lattice\cite{dal00}. At the same time, there is no clear view how to extrapolate results into the small region $a$ (see the remark in Ref.~\cite{dal99}.)
Also, a "naive" LF Hamiltonian is used usually.
That is why new interesting results may be obtained if we construct a "corrected" LF Hamiltonian
corresponding to renormalized theory with non-physical degrees of freedom switching off in the limit $a \to 0$.
This can be useful for a substantiation of transverse lattice method.

In present paper, we take a first step to the construction of a "corrected" LF Hamiltonian for a theory with transverse lattice. Instead of QCD, we consider a pure nonabelian gauge field theory. Transition from such model to full QCD is not so complicated and it is fully based on the results, obtained for this model.
Considering pure nonabelian gauge theory allows to focus on the features of this theory.

We propose the action of theory on the lattice (Sect.~2),
construct Feynman perturbation theory (Sect.~3),
and consider all Feynman diagrams that do not contain divergences (Sect.~5). We can show that in the some limits $a \to 0$ and $m \to \infty $ such diagrams are either equal to corresponding diagrams of continuous theory, either equal to zero (which corresponds to switching off the non-physical degrees of freedom).
The obtained result is also true for diagrams with divergence,
but only after applying some procedure of
subtraction, if in a result of this procedure the indices of ultraviolet (UV)
divergence  $\om,\om_\p$ become negative.
This fact is very significant for renormalization of the theory, because it is the main tool at
the analysis of divergent parts of diagrams.
It is necessary to note, that limits $a \to 0$ and $m \to \infty $ are not independent.
Parameter $m$ plays a role of mass for non-physical degrees of freedom and
the above formulated statement is true only if $m$ a function of $a$ satisfying certain conditions.

The statement  mentioned above is enough for renormalization of the theory on transverse lattice if we do not take into account the  possibility of occurrence of non-polynomial with respect to momenta divergences.
Such possibility exists because Lorentz invariance is broken in the given theory, however the results of preliminary analysis show that non-polynomial divergences are absent for Green functions.
In this paper we do not discuss this subject, we do not go beyond the description of the scheme
of renormalization procedure (Sect.~6).
The accurate construction of this procedure which leads to occurrence of action of the theory with all necessary counterterms will be presented in the further work.
In Sect.~4 we discuss
the problems of UV divergences of the theory that remain when transverse lattice is applied.

It is necessary to notice that in this paper we use gauge group $U(N)$ instead of $SU(N)$.
This fact is not significant because for pure nonabelian gauge field theory, abelian and nonabelian components of gauge field do not interact in the limit $a\to0$. We do not use immediately the group $SU(N)$, because for $SU(N)$ the propagator of the non-physical degree of freedom, corresponding to abelian part of transverse gauge field, do not sufficiently decrease at large momenta.

\section{Action theory on transverse lattice}
We formulate gauge theory in 4D space-time, where two space-like directions are replaced with the square lattice. Following Refs.~\cite{barpir1,heplat,tmf04}, we choose variables of the theory in such a way that the actions of the theory are polynomial.
In this case, the components of gauge field along continuous coordinates are described in a usual way and related to the vertices of the lattice.
The components of field along discrete coordinates $x^k$, $k,l,\ldots=1,2$ are described by complex $N\times N$ matrices (in the case of $U(N)$ theory.) The matrix $M_k(x)$ relates to the link of lattice that connects the vertices $x-e_k$ and $x$, and corresponds to positive direction along axis $x^k$, see Fig.~1, {\it a}.
\begin{figure}[htb]
\centerline{\psfig{file=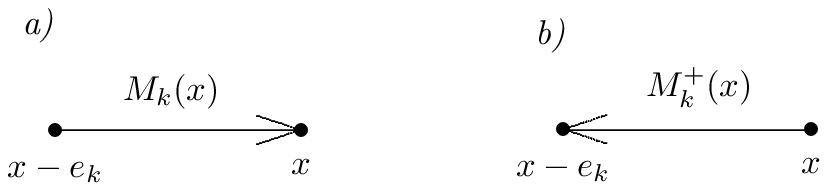,height=5em}}
\caption{}
\end{figure}
The vector $e_k$ connects neighbor vertices of lattice and is directed along the axis $x^k$, so that $|e_k|=a$, where $a$ is the lattice spacing. Matrices $M_k(x)$ are considered as independent variables. These matrices are arbitrary (non-unitary), and actions in these variables can be chosen as polynomial. The matrix $M_k^+(x)$ is related to the same link as $M_k(x)$, but corresponds to the opposite direction, see fig.~1, {\it b}.

Trace of matrices' product may be related to any closed directed cycle on the lattice, if the matrices are placed on the links and are directed along the cycle. For example, the expression
\disn{1}{
{\rm Tr}\left\{M_2(x)M_1(x-e_{2})M_2^+(x-e_{1})M_1^+(x)\right\}
\nom}
corresponds to the cycle shown in Fig.~2, {\it a}.
\begin{figure}[htb]
\centerline{\psfig{file=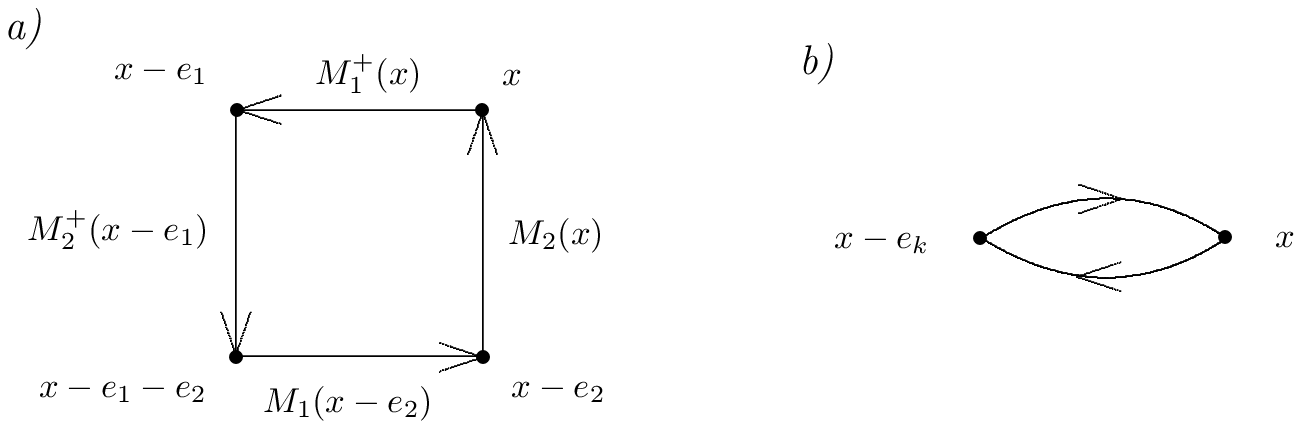,height=11em}}
\caption{}
\end{figure}
It should be noted that the trace related to the closed cycle consisting of the same links passed in both directions, does not correspond to unity because the matrices W are not unitary.
(see. Fig.~2, {\it b})

Gauge transformations given by unitary $N\times N$ matrices $U(x)$ are applied, as usual, to longitudinal components of the field $A_C(x)$:
\disn{2p}{
A'_C(x)=U(x)A_C(x)U^+(x)+\frac{i}{g}U(x)\dd_C(x)U^+(x).
\nom}
They are applied to the matrices $M_k(x)$ as:
 \disn{2}{
M'_k(x)=U(x)M_k(x)U^+(x-e_{k}).
\nom}
Any trace of matrices that corresponds to the cycle (for example, Fig.~2) is invariant of these transformations.

We find the relation between the matrix variables $M_k(x)$ and the gauge field $A_k(x)$ which arise in the continuous limit $a \to 0$.
The field $A_k(x)$ consists of abelian and nonabelian parts corresponding to group $SU(N)$ and $U(1)$. Namely:
\disn{3}{
M_k(x)=I+ga\ls B_k(x)+iA_k(x)\rs\equiv I+ga\,V_k(x),
\nom}
where $I$ is a unit matrix, $A_k(x)$, $B_k(x)$ are Hermitian matrices.
Field $B_k(x)$ is an auxiliary non-physical field which should be switched off in the limit $a \to 0$.

For the lattice model, the transverse component of field strength $G_{\m\n}$ can be determined by different ways. Pure longitudinal components of $G_{CD}$ correspond to the vertex of lattice and have usual form:
\disn{4}{
G_{CD}(x)=\dd_{C}A_{D}(x)-\dd_{C}A_{D}(x)-ig[A_{C}(x),A_{D}(x)].
\nom}
We determine mixed components of $G_{Ck}$ as:
\disn{5}{
G_{C,k}(x)=\frac{1}{iga}\ls \dd_{C}M_k(x)-ig\ls A_{C}(x)M_k(x)-
M_k(x)A_{C}(x-e_k)\rs\rs,
\nom}
they correspond to the same link as $M_k(x)$. Gauge transformations to $G_{Ck}$ are similar to (\ref{2}).

For pure transverse intensity components $G_{kl}$ we use two representations.
The first representation is:
\disn{6}{
G^{(1)}_{kl}(x)=\frac{i}{ga^2}\Bigl(M_k(x)M_l(x-e_k)-M_l(x)M_k(x-e_l)\Bigr)
 =\ns=
\frac{i}{ga^2}\Biggl(
\raisebox{-1.6em}{\psfig{file=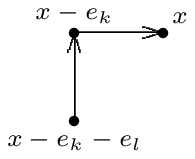,height=4em}}
-
\raisebox{-1.7em}{\psfig{file=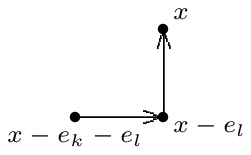,height=4em}}
\Biggr)
\nom}
which is antisymmetrical with respect to permutation $k,l$ and is transformed as:
 \disn{6.1}{
{G^{(1)}_{kl}}'(x)=U(x)G^{(1)}_{kl}(x)U^+(x-e_{k}-e_{l}).
\nom}
The second representation is:
 \disn{7}{
G^{(2)}_{kl}(x)=\frac{i}{ga^2}\Bigl(M_l(x-e_k)M^+_k(x-e_l)-M^+_k(x)M_l(x)\Bigr),
\nom}
or
 \disn{8}{
G^{(2)}_{kl}(x)=
\frac{i}{ga^2}\Biggl(
\raisebox{-1.6em}{\psfig{file=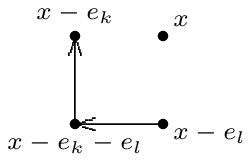,height=4em}}
-
\raisebox{-1.6em}{\psfig{file=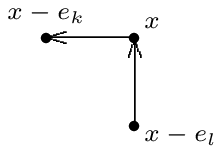,height=4em}}
\Biggr),\quad k\ne l,\no
G^{(2)}_{kk}(x)=\frac{i}{ga^2}
\Biggl(\hskip 0.5em
\raisebox{-0.6em}{\psfig{file=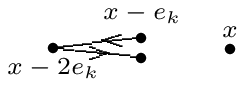,height=2.1em}}
\hskip 1em - \hskip 1em
\raisebox{-0.1em}{\psfig{file=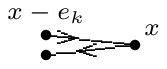,height=1.6em}}
 \Biggr).
\nom}
This representation is not antisymmetric (it satisfies to the condition
$G^{(2)}_{lk}=-G^{(2)+}_{kl}$) and it is transformed as:
\disn{9}{
{G^{(2)}_{kl}}'(x)=U(x-e_{k})G^{(2)}_{kl}(x)U^+(x-e_{l}).
\nom}

We take the following expression as starting action of the theory:
 \disn{10}{
S^M=a^2\sum_{x^\p}\;\int\! d^2 x^{\pa}\ls L_1+L_2+L_3+L_m\rs,
\nom}
where
 \disn{11}{
L_1=-\frac{1}{2}\,\sum_{C,D}\tr\ls G_{CD}G^{CD}\rs=\tr\ls G_{03}G_{03}\rs,\no
L_2=-\sum_{C,k}\tr\ls G^+_{Ck}G^{Ck}\rs,\no
L_3=-\frac{1}{4}\,\sum_{k,l}\tr\ls G^{(1)+}_{kl}G^{(1)}_{kl}+G^{(2)+}_{kl}G^{(2)}_{kl}\rs,\no
L_m=-\frac{m^2}{4g^2a^2}\,\sum_{k}\tr\ls\ls M_k(x)M_k^+(x)-I\rs^2\rs.
\nom}
This action was suggested in the work Ref.~\cite{heplat}.

Lets consider then the symmetry of action (\ref{10}). Obviously, each term of the action is invariant of gauge transformation.
Complete Lorentz invariance was broken by the transverse lattice. Instead of Lorentz invariance we have the Lorentz boost invariance in plane of the continuous coordinates $x^0,x^3$.
Also, we have an invariance relatively to the discrete group of rotations by $\pi/2$ in plane $x^1,x^2$.
This can be proved by following. The contribution of quantity $L_1$ is the sum over all vertices, and the contributions of quantities $L_2$ and $L_m$ are the sums over all links (if we take into account the summation over $k$).
Quantity $L_3$ can be written as
 \disn{12}{
L_3=-\frac{1}{2g^2a^4}\Biggl(
\raisebox{-1.6em}{\psfig{file=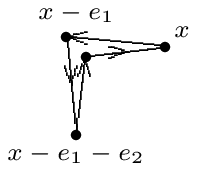,height=4.5em}}
+
\raisebox{-1.8em}{\psfig{file=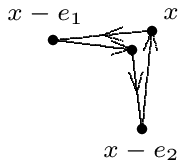,height=4.5em}}
+
\raisebox{-1.6em}{\psfig{file=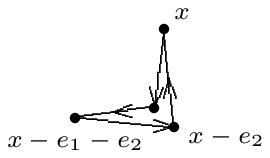,height=4.5em}}
+
\raisebox{-1.6em}{\psfig{file=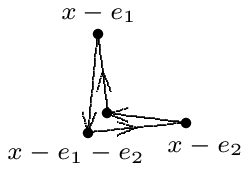,height=4.5em}}
-\ns-2\Biggl(
\raisebox{-1.6em}{\psfig{file=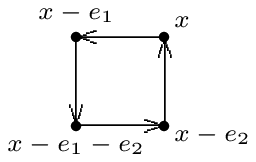,height=4.5em}}
+
\raisebox{-1.6em}{\psfig{file=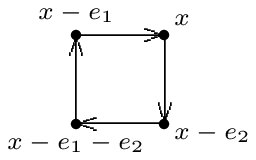,height=4.5em}}
 \Biggr)\Biggr)-\no
-\frac{1}{4g^2a^4}\sum_k\Biggl(
\raisebox{-0.6em}{\psfig{file=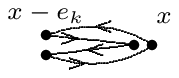,height=2.2em}}
+
\raisebox{-0.6em}{\psfig{file=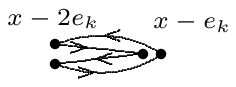,height=2.2em}}
-2\;
\raisebox{-0.6em}{\psfig{file=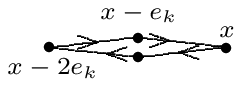,height=2.4em}}
\Biggr),
\nom}
Invariance of this expression is obvious after summation along $x^\p$.

It is useful to formulate the transformation rule for the fields under the action of discrete group
of rotations by $\pi/2$ in plane $x^1,x^2$.
Obviously, if we introduce new coordinates $x'$ from old coordinates $x$ by rotating the axis $1,2$ on $\pi/2$, we obtain relations:
 \disn{12.1}{
{x^1}'=-x^2,\quad {x^2}'=x^1,\qquad {e_1}'=-e_2,\quad {e_2}'=e_1,\no
M'_1(x')=M_2^+(x+e_2),\qquad M'_2(x')=M_1(x).
\nom}
Hence, taking into account (\ref{3}), we can find the rule of transformation:
 \disn{12.2}{
A'_1(x')=-A_2(x+e_2),\quad A'_2(x')=A_1(x),\no
B'_1(x')=B_2(x+e_2),\quad B'_2(x')=B_1(x).
\nom}
This equations will be used in later.
It is worth noting that in the limit $a\to 0$ field $A_k$
transforms as a usual vector and, in particular, changes sign when rotated by $\pi/2$.
The transformation rule for the field $B_k$ is different and the field $B_k$ is invariable by rotation by $\pi/2$.

Let us consider the action (\ref{10}) in the "naive" limit $a\to 0$. At first, we consider expressions (\ref{4}),(\ref{5}),(\ref{6}),(\ref{7}) in the limit $a\to 0$. If we introduce $F_{\m\n}$ as:
 \disn{13}{
F_{\m\n}=\dd_{\m}A_{\n}-\dd_{\n}A_{\m}-ig[A_{\m},A_{\n}],
 \nom}
then we have:
 \disn{13.1}{
G_{CD}=F_{CD},\qquad
G_{Ck}\str{a\to0}\hat G_{Ck}=F_{Ck}-iD_CB_k,\no
G^{(1)}_{kl}\str{a\to0}\hat G^{(1)}_{kl}=F_{kl}-i\ls D_kB_l-D_lB_k\rs+ig[B_k,B_l],\no
G^{(2)}_{kl}\str{a\to0}\hat G^{(2)}_{kl}=F_{kl}-i\ls D_kB_l+D_lB_k\rs-ig[B_k,B_l],
\nom}
where
 \disn{13.2}{
D_\m B_k(x)=\dd_\m B_k(x)-ig[A_\m(x),B_k(x)].
\nom}
It should be noted that the fields $A_\m$ and $B_k$ contain nonabelian and abelian components.

We consider $L_m$ in the limit $a\to 0$ at fixed parameter $m$. We have
 \disn{14}{
L_m\str{a\to0}\hat L_m=-m^2\sum_k\tr\ls B_kB_k\rs.
\nom}
Using the expressions in (\ref{10}),(\ref{11}) we have:
 \disn{15}{
S^M\str{a\to0}\int d^4x\;\tr\ls-\frac{1}{2}F_{\m\n}F^{\m\n}+(D_\m B_k)(D^\m B_k)-g^2[B_k,B_l][B_k,B_l]-m^2B_kB_k\rs.
\nom}
Here and further, we follow the convention of the summation over repeated twice indices.

With the form of "naive" limit of action (\ref{15}) we expect that non-physical degree of freedom $B_k$
is switched off it the limit $a\to0$ and $m\to\infty$. It is easy to see that taking into account (\ref{15}), abelian
and nonabelian parts of field $A_\m$ do not interact with each other and their contribution to
actions are standard. Hence it is justified to expect that Green functions of field $A_\m$ of this theory tend to
Green functions of standard nonabelian and abelian gauge field in continuous limit.
For verifying this fact, it is necessary to construct Feynman rules of theory on transverse lattice with additional field $B_k$ and to renormalize the theory.

\section{Feynman perturbation theory on transverse lattice}
We construct the Feynman perturbation theory in space-time with transverse lattice.
For this, we introduce certain notations and determine the discrete derivative along the axes $x^1,x^2$ as:
 \disn{16}{
\ddl_k\ff(x)=\frac{1}{a}\ls \ff(x)-\ff(x-e_k)\rs,\qquad
\ddr_k\ff(x)=\frac{1}{a}\ls \ff(x+e_k)-\ff(x)\rs.
\nom}
Both expressions become standard derivatives in the limit $a \to 0$.
The equations
 \disn{17}{
[\ddl_i,\ddl_k]=[\ddr_i,\ddr_k]=[\ddl_i,\ddr_k]=0,
\nom}
are true. The equation:
 \disn{18}{
\sum_{x^\p}\ls\ddl_k f(x)\rs g(x)=-\sum_{x^\p}f(x)\ls\ddr_k g(x)\rs
\nom}
gives the discrete analogue to partial integration and is true as well.

The momentum space along the
directions $k_1,k_2$ is finite for the fields
determined in
coordinate space with the transverse lattice. Size of this space is $2\pi/a$.
The fields in momentum space are periodical functions.
The relations between functions $\ff(x)$ and $\ff(k)$ are given by
 \disn{18.1}{
\ff(k)=a^2\sum_{x^\p}\int d^2x^{\pa}e^{-ik_\m x^\m}\ff(x),\qquad
\ff\ls k+\frac{2\pi}{a}\frac{e_l}{a}\rs=\ff(k),\no
\ff(x)=\frac{1}{(2\pi)^4}\int_{-\pi/a}^{\pi/a}dk_1\int_{-\pi/a}^{\pi/a}dk_2\int d^2k_{\pa}
e^{ik_\m x^\m}\ff(k),
\nom}
which means that certain Fourier integrals are replaced with Fourier
series.
Here and below we assume that the sign changes when lower indices $k,l,\ldots$ are lifted, i.e. $k_\m x^\m=k_Cx^C+k_lx^l=k_Cx^C-k_lx_l$.

As we can see from (\ref{18.1}), in momentum space, the discrete derivatives (\ref{16}) become
operators of products:
 \disn{18.2}{
\ddl_l\to \frac{1}{a}\ls 1-e^{-ik_la}\rs\equiv iu_l,\qquad
\ddr_l\to \frac{1}{a}\ls e^{ik_la}-1\rs=iu^*_l,
\nom}
where "$*$" is complex conjugation.
If we denote $u_C\equiv k_C$, then we can write $\ddl_\m\to iu_\m$. We note that:
 \disn{18.3}{
u_l=e^{-ik_la/2}\;\,\frac{\sin(k_la/2)}{a/2}\;\str{a\to 0}\;k_l.
\nom}
We also introduce notations:
 \disn{18.4}{
|u_\p|^2\equiv u^*_l u_l=\sum_l\ls\frac{\sin(k_la/2)}{a/2}\rs^2\str{a\to 0}k_\p^2,\qquad
|u|^2\equiv k_{\pa}^2-u^*_l u_l\str{a\to 0}k^2.
\nom}

Then we can construct the Feynman perturbation theory by $g$.
One can use fields $A_\m$ and $B_k$ as independent variables, because they are equivalent to variables $A_C(x)$ and complex matrix fields  $M_k(x)$ according to (\ref{3}).
We can rewrite the action (\ref{10}) in terms
of fields $A_\m$ and $B_k$.
For that, we rewrite equations (\ref{5}),(\ref{6}),(\ref{7}) with discrete derivatives.

For example, the equation (\ref{5}) can be written as:
 \disn{19}{
G_{Ck}(x)=\dd_{C}A_{k}(x)-\ddl_{k}A_{C}(x)-ig[A_{C}(x),A_{k}(x)]-iD_CB_k(x)-ga V_k(x)\ddl_k A_C(x),
\nom}
where $V_k(x)$ is defined by the equation (\ref{3}).
Obviously, this equation differs from the limit value $\hat G_{C,k}(x)$ (\ref{13.1}) (where $\dd_k$ should be understood as $\ddl_k$), but the difference is in the last item only. This item has an extra factor $a$ and vanishes in the limit $a \to 0$.
In our further calculations only the structure of such
expressions (not their precise value) will be important. Therefore we will write these equations
symbolically and omit the numerical factors and indices. Moreover, any linear combination of fields $A_\m$, $B_k$ will be written as $V$. Then the formula (\ref{19}) can be outlined as:
 \disn{20}{
G_{Ck}(x)=\hat G_{Ck}(x)+gaV\ddl_\p A_{\pa},
\nom}
where symbol ${\scriptstyle \|}$ corresponds to the value of indices $0,3$ and $\p$ corresponds to the value of indices $1,2$. Similar,
 \disn{21}{
G^{(1)}_{kl}(x)=\hat G^{(1)}_{kl}(x)+gaV\ddl_\p V,\no
G^{(2)}_{kl}(x)=\hat G^{(2)}_{kl}(x)+gaV\ddl_\p V+ga^2\ls\ddl_\p V\rs^2,
\nom}
where we assume $\ddl_k$ instead of $\dd_k$
in quantities with hat.
Then, the quantity of $L_m$ can be written as:
 \disn{21.1}{
L_m=\hat L_m+
m^2\Bigl( ga\,\tr\ls BV^2\rs+g^2a^2\tr\ls V^4\rs\Bigr).
\nom}

Using formulas
 (\ref{20})-(\ref{21.1}), action (\ref{10}) can be expressed as:
 \disn{22}{
S^M=a^2\sum_{x^\p}\!\int\! d^2 x^{\pa}
\biggl(\tr\ls-\frac{1}{2}\tilde F_{\m\n}\tilde F^{\m\n}+(\tilde D_\m B_k)(\tilde D^\m B_k)-g^2[B_k,B_l][B_k,B_l]-m^2B_kB_k\rs+\ns
+ga\,\tr\ls V(\dd_{\pa} V)(\ddl_\p A_{\pa})\rs+
\sum_{\xi,\de}g^{\xi+\de-2}a^{\xi+2\de-4}\tr\ls V^\xi(\ddl_\p V)^\de\rs+
\ns+
m^2 ga\,\tr\ls BV^2\rs+m^2 g^2a^2\tr\ls V^4\rs
\biggr),
\nom}
where pair of indices
$\xi,\de$ can have values $(0,3),(0,4),(1,2),(1,3),(2,2),(3,1)$.
We imply summation over repeated index. Values $\tilde\dd_\m$, $\tilde D_\m$, $\tilde F_{\m\n}$
correspond to the value with discrete derivatives $\ddl_k$ instead of standard derivatives on transverse plane $\dd_k$.

It is necessary to fix the gauge
for further development of perturbation theory. We use light-like gauge $A_-=0$ because
in future we will use this perturbation theory for constructing "corrected" LF
Hamiltonian (see Introduction).
This gauge does not lead to appearance of Faddeev-Popov ghosts.
We can introduce this gauge on transverse lattice by adding the term
 \disn{23}{
S^{{\rm gf}}=-2a^2\sum_{x^\p}\!\int\! d^2 x^{\pa}\,\tr\ls\La\, n^\m A_\m\rs
\nom}
to the action (\ref{22}), where $n^\m$ is the light-like vector lying in the coordinates plane $x^C$, $n^-=1$, $n^{+,k}=0$, and $\La$ is additional Hermitian matrix field considered as an additional independent variable.

We select a part $S_0$, which is free and quadratic by fields $A_\m$, $B_k$, from the action $S=S^M+S^{{\rm gf}}$:
 \disn{24}{
S_0=a^2\sum_{x^\p}\!\int\! d^2 x^{\pa}
\;\frac{1}{2}\biggl(-(\ddl_\m A^a_\n)(\ddl^\m A^{a\n})+(\ddl_\m A^{a\m})^2-2\La^a n^\m A^a_\m+\ns+
(\ddl_\m B^a_k)(\ddl^\m B^a_k)-m^2B^a_kB^a_k\biggr),
\nom}
where $A^a_\m,B^a_\m,\La^a$ are coefficients of decomposition of corresponding matrices by the basis
of Hermitian matrices $\la^a/2$,\ \ $a=0,1,2,\ldots$. The properties of these Hermitian matrixes
are: $\la^0=\sqrt{2/N}\,I$, $\tr\,(\la^1)=\tr\,(\la^2)=\ldots=0$,
$\tr\ls\la^a\la^b\rs=2\de^{ab}$, $[\la^a,\la^b]=2if^{abc}\la^c$, where $f^{abc}$ are structure
constants, $f^{0bc}=0$.
We can obtain propagators after partial integration using (\ref{18}),
finding of inverse quadratic form and making Fourier transformation using (\ref{18.1}).
If we use formulas (\ref{18.2}) and (\ref{18.4}) the propagators are obtained as
 \disn{25.1}{
\De^{(AA)ab}_{\m\n}(k)=-\frac{i\de^{ab}}{|u|^2+i0}\ls g_{\m\n}-\frac{u^*_\m n_\n+n_\m u_\n}{k_{\pa}^2+i0}\,2k_\al \bar n^\al\rs,
\nom}
\vskip -1em
 \disn{25.2}{
\De^{(A\La)ab}_{\m}(k)=-\frac{i\de^{ab} u_\m}{k_{\pa}^2+i0}\,2k_\al \bar n^\al,\qquad
\De^{(\La\La)ab}=0,
\nom}
\vskip -1em
 \disn{25.3}{
\De^{(BB)ab}_{lm}(k)=\frac{i\de^{ab}\de_{lm}}{|u|^2-m^2+i0},
\nom}
where $\bar n^+=1$, $\bar n^{-,k}=0$, and we use the Mandelstam-Leibrandt\cite{man,leib} prescription
for poles.
With this prescription we can make transition to
Euclidean space. We note that propagator
$\De^{(\La\La)ab}$ of the field $\La^a$ with itself equals zero.
However, non-diagonal propagator
$\De^{(\La\La)ab}$ of field $\La^a$ with $A_\m^a$ does not equal zero. All other non-diagonal
propagators are equal zero.

The result of subtracting free part $S_0$ from action $S$ is the action of interaction,
the items of which give vertices of Feynman diagrams. We call the vertices corresponding
to four last items in
(\ref{22}) "extra vertices".
These items has an additional factor $a$ which vanishes in the limit $a\to0$.
The action of interaction contains usual terms
(in which $\dd_\m$ is replaced on $\ddl_\m$) which are the terms
of third- and fourth-order self interaction of field $A^a_\m$:
 \disn{26}{
S^1_I=a^2\sum_{x^\p}\!\int\! d^2 x^{\pa}\ls
-gf^{abc}A^a_\m A^b_\n \ddl^\m A^{c\n}-\frac{g^2}{4}
f^{abe}f^{cde}A^a_\m A^b_\n A^{c\m} A^{d\n}\rs,
\nom}
and terms of interaction of fields $A^a_\m$ with $B^a_k$ and of self interaction of field $B^a_k$:
 \disn{27}{
S^2_I=a^2\sum_{x^\p}\!\int\! d^2 x^{\pa}\ls
gf^{abc}A^a_\m B^b_k \ddl^\m B^c_k+\frac{g^2}{2}f^{abe}f^{cde}
\ls A^a_\m B^b_k A^{c\m} B^d_k+B^a_k B^b_l B^c_k B^d_l\rs\rs.
\nom}
We note that since $f^{0bc}=0$, the abelian components of fields are not included in these
expressions. That is why,
these components of fields interact with other fields via "extra" vertices only.

\section{Longitudinal ultraviolet divergences}
In this chapter, we describe an approach which allows to use an analogue of Word identities,
despite the fact that the transverse lattice does not give
full UV regularization of the theory.

At first we would like to identify UV divergences which exist in the theory. Obviously, if step of lattice $a$ is
finite, then Feynman integral can diverge on longitudinal components $k_0,k_3$ of momentum only.
The interval of integration along transverse component of momentum $k_1,k_2$ is
finite.
Therefore, for the case of
nonzero $a$, the only diagrams that diverge are the
one-particle irreducible (1PI) diagrams (or diagrams containing such subdiagrams)
having $\om_{\pa}\ge0$, where $\om_{\pa}$ is
the index of UV divergence on subspace $k_0,k_3$.
We can find all such
1PI diagrams if
we consider contributions from
propagators (\ref{25.1})-(\ref{25.3}) and vertices,
defined by (\ref{26}), (\ref{27}) and
by three last terms of formula (\ref{22}).
Such diagrams include all diagrams with one vertex and
shorted lines, and also include the diagram
shown in Fig.~1,
 \begin{figure}[htb]
\centerline{\psfig{file=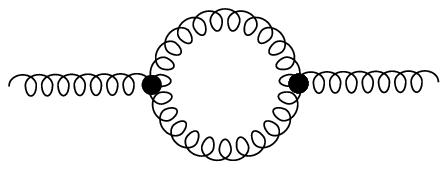,height=3.5em}}
\caption{}
\end{figure}
where lines correspond to fields $A_\m$ or $B_k$.

Thus, the introduction of transverse lattice does not lead to total
UV regularization of this theory. However, only
finite number of diagrams are UV divergent.
For the development of the perturbation theory
it is necessary to introduce additional regularization in order to make theory totally finite.
Unfortunately, we do not know a way to introduce an additional regularization if we want to use the method of
canonical quantization on Light Front and save gauge invariance.
If we permit to break gauge invariance, then
we can take the Hamiltonian of interaction in normal ordered form.
In this case all diagrams with
shorted lines vanish. Also, we can add hight non-covariant derivative
with regularization parameter $\La$ (like in paper Ref.~\cite{tmf99}) in action
for the regularization of diagrams like the one shown on Fig.~3.
We call it the "first" form of theory. The first form of theory is not gauge invariant.
That is why, analogues of Word identities can not be formulated for this theory.

We also consider the so called "second" form of theory. Additional regularization of this theory
is the dimensional regularization with parameter $\e$. It is well known, that the dimension regularization
preserves gauge invariance.
We can introduce the dimensional regularization for the theory with
transverse lattice by the method which is described in book Ref.~\cite{col}.
According to this method, integration along transverse direction should be performed
only after integration along longitudinal direction. The square of
longitudinal part of momentum
after Wick rotation
should be continued up to the square of the infinite-dimensional vector.

Consider the limit $\e\to0$ for the second form of theory. We
will calculate diagrams
similar to the one in Fig. 3 and diagrams with single vertex and shorted lines
in dimensional regularization. After that, we will obtain "renormalized"
(in limit $\e\to0$, but at fixed $a$) values of these
diagrams using the scheme of minimal subtraction,
i.e.~discarding poles at $\e$.
For other 1PI diagrams we can take limit $\e\to0$.
Note, that this limit is coincide with the limit $\La\to\infty$ for
corresponding diagram of first form of theory.
As a result, we will obtain value for all diagrams in the limit
$\e\to0$, we will call it the "third" form of theory.
By construction, an analogue of Word identities can be constructed for this third form of theory, which makes it useful for analysis of divergence in limit $a\to0$.

It is necessary to add counterterms to action of first form of theory to compensate the divergence of diagram in Fig.~3
in the limit $\La\to\infty$ so
that this diagram will coincides with "renormalized" values of diagram of third form of theory calculated
by dimensional regularization. These counterterms do not contain derivatives, because for
such diagrams $\om_{\pa}=0$. If we apply
such procedure to diagrams with one vertex and
shorted lines, then
the  counterterms should only contain
finite (in $\La\to\infty$ limit) part, because these
diagrams are equal to zero in the first form of the theory.
Note, that we have a finite (and not very large) number of diagrams
which require the application of described procedure.
As a result, all Green functions of the first form of the theory (including the added counterterms)
in the limit $\La\to\infty$ will be equal to corresponding Green functions of the third form of the theory.

Later in this work, we will study only the third form of the theory,
therefore we  will assume that longitudinal
divergence is eliminated via gauge invariant way. It is necessary to show,
first of all,
that its Green functions are equal (after renormalization) to the Green functions
of renormalized theory in continuous space and that the non-physical fields are switched off.

After that, we should correctly formulate the first form of theory choosing
some gauge
noninvariant longitudinal
regularization and calculating necessary diagrams (see above).
Green functions of this theory (including the added counterterms) in the limit $\La\to\infty$
(at first),
and $a\to0$ will coincide with renormalized Green functions of continuous theory.
Just the first form of the theory can be used for construction of
"corrected" LF Hamiltonian
(see Introduction) and for nonperturbative calculations.
The exact construction of this theory is beyond the aim of this paper.

\section{Analysis of non-divergent diagrams}
Let us show that all diagrams not containing UV divergence and with external lines of type $A^a_\m$
either coincide with corresponding diagrams of continuous theory either vanish
in the limit $a\to0$ (and in the limit $m\to\infty$).
This corresponds to the switching off of the non-physical degrees of freedom.

Consider 1PI diagrams of Feynman perturbation theory with propagators and terms of interaction
described in Sect.~3. Lets take into account the Wick rotation $k_0=ik_4$, i.~e.~calculate
diagrams in Euclid space.
We note that if we perform transition in Euclid space, then vector $n^\m$, concerned with the gauge condition,
becomes a complex vector $n^4=i/\sqrt{2}$, $n^3=-1/\sqrt{2}$, while the vector $\bar n^\m$,
which is contained in propagators (\ref{25.1}),(\ref{25.2}) becomes the vector $\bar n^\m={n^\m}^*$
which is a complex conjugate to $n^\m$.

The index of UV divergence $\om$ plays a significant role for analysis of diagrams.
In this theory $\om$ should be defined in the non-standard way
because of extra vertices and unusual propagators. We consider an arbitrary multiloop 1PI diagram and write
it symbolically as:
 \disn{28}{
D=\int\! dk\; F(k).
\nom}
Here, $k$ denotes all momenta of integration.
We now find the main term of integrands expression $F(k)$ at $a\to0$ at fixed $m$.
It has the form $m^{\ga_m}a^{\ga_a}\tilde F(k)$, where $\tilde F(k)$ is a usual Feynman integrands
expression (as a consequence of (\ref{18.3}),(\ref{18.4})), while $\ga_m$ and $\ga_a$ are certain
non-negative numbers which depend on the diagram type only (and determined by number and type
of "extra" vertices).
We assume that parameter $m$ increases as $1/a$ with accuracy up to logarithmic corrections in the limit $a\to0$.
One can show that without this assumption the switching off of the non-physical degrees of freedom
does not take place.

We use the notation $\om'$ to denote the usual index of UV divergence of integral $\int\! dk\,\tilde F(k)$.
We call the value of
 \disn{29}{
\om=\om'+\ga_m-\ga_a
\nom}
as the generalized index of UV divergence of the diagram.
It will be seen later that this value determines the convergence of diagrams
(we remind that value $\pi/a$ plays the role of transverse momenta cutoff).
We now introduce the generalized index of divergence in transverse direction:
 \disn{29.1}{
\om_\p=\om'_\p+\ga_m-\ga_a.
\nom}
It can be useful because Lorentz invariance is broken and divergence in transverse direction $k_1,k_2$ may be stronger than the total divergence.

Consider behavior of diagrams in the limit $a\to0$ and take into account the behavior of $m(a)$.
Limit ourself in this section only those diagrams that have external lines, corresponding to field $A^a_\m$,
and generalized indices $\om,\om_\p<0$ (as well as longitudinal index $\om_{\pa}<0$ too).
This fact should be true
for total diagram and for all subdiagrams of it as well.
It may be noted that if $\om,\om_\p<0$, then $\om,\om_\p\le -1$.

As the first step, we consider diagrams without propagators of field $B^a_k$. Note that
the set of these diagrams is equivalent to the
set of diagrams corresponding to continuous gauge field theory with group $U(N)$.
For these diagrams $\ga_a=\ga_m=0$, which leads to $\om'=\om<0$ and $\om'_\p=\om_\p<0$. The integrand expressions
$F(k)$ for these diagrams are the products of propagators (\ref{25.1}) and vertex factors
like $u_\m$ (see definition in eqs.~(\ref{18.2}),(\ref{18.3}) and in text between) from (\ref{26}).

We split integral (\ref{28}) into two terms
 \disn{30}{
D=\int\! dk\; F(k)=\int\! dk\;\tilde F(k)+\int\! dk\;\ls F(k)-\tilde F(k)\rs,
\nom}
where $\tilde F(k)$ is $F(k)$ in the limit $a\to 0$.
The parameter $a$ vanishes from the integrands expression of the first term which is identical to the standard integrands
expression, corresponding to the diagram of continuous theory.
However, $a$ remains in the limit of integration of the first term.
The momenta of each line are limited by condition
$|k_{1,2}|\le \pi/a$.
As long as $\om',\om'_\p,\om_{\pa}<0$ (for total diagram and for all subdiagrams of it)
this term is finite in the limit $a\to 0$ and is identical to the result
of calculation of corresponding diagram of continuous theory.

The second item of (\ref{30}) vanishes in the limit $a\to 0$. Let's show this fact.
The values $F(k)$ and $\tilde F(k)$ are fractions, hence we can bring their difference to a common denominator.
The numerator of this expression can be presented as a sum
of terms with factors $(u_l-k_l)^n$, $n\ge1$.
The momenta of lines are limited as $|k_{1,2}|\le \pi/a$,
therefore we can use the estimation:
 \disn{31}{
|u_l-k_l|^n\le\ls\frac{1}{2} a\, k_l^2\rs^n.
\nom}
We can use also the estimation
 \disn{32}{
|u_l|\le |k_l|,
\nom}
for other quantities $u_l$ in numerator, and the estimation
 \disn{33}{
|u_\p|^2 \ge \frac{4}{\pi^2}\, k_\p^2
\nom}
for quantities in denominator.
In a result we can estimate that
 \disn{34}{
\left|\int\! dk\;\ls F(k)-\tilde F(k)\rs\right|\le
a^n\int\! dk\;\hat F(k),
\nom}
where quantity $\hat F(k)$ does not contain factor $a$ and has a usual Feynman form.
The UV indices $\om',\om'_\p$ for integral in right-hand part of estimation (\ref{34})
are larger than the corresponding indices of initial diagram by $n$.
The increase of indices
and appearance of factors $a^n$ is a consequence of using of the estimation (\ref{31}).
The indices of the initial diagram are negative. Therefore the indices of integral in right-hand part of estimation (\ref{34}) are less than or equal to $n-1$.
Then, this integral can be estimated as $1/a^{n-1}$ in the limit $a\to0$. The selected factor $a^n$ causes the left-hand part of (\ref{34}) to vanish in the limit $a\to0$.

As a result, the value of diagrams of the considered class, without propagators of field $B^a_k$
and without extra vertices, tend (in the limit $a\to0$) to the values of corresponding diagrams of continuous gauge field theory with group $U(N)$.

As the second step, we consider other diagrams of considered class,
i.e. diagrams with propagators of field $B^a_k$ and (or) extra vertices. These diagrams vanish in the limit $a\to0$.
Let us show this fact. If a diagram contains the propagators of field $B^a_k$ (we assume the number
of such propagators to be equal to $\be\ge0$), then we
can estimate euclidian form of these propagators
 \disn{35}{
\frac{1}{|u|^2+m^2}\le \frac{1}{m^2}.
\nom}
We can estimate the rest parts of integrand in eq.~(\ref{28})
similarly as above. The quantities $u_l$ in the numerator can be
estimated by (\ref{32}) and
the quantities $|u_\p|^2$ in
the denominator can be estimated by (\ref{33}). As a result, the value of diagram can be estimated as
 \disn{36}{
m^{\ga_m-2\be}a^{\ga_a}\int\! dk\;\hat F(k),
\nom}
where $\hat F(k)$, similarly as above, has a usual Feynman form.

We now analyse the form of extended vertices in eq.~(\ref{22}).
Obviously, the difference $\ga_m-\ga_a$ is less than or equal to the number of vertices
which correspond to the
penultimate term of (\ref{22}) (the contribution of
such a vertex in $\ga_m-\ga_a$ is equal to 1, contribution of other extended vertices is less than or equal to zero).
Mentioned vertex should be connected to one or more propagators of field  $B^a_k$. As a result,
a number of these vertices is less than or equal to the doubled number of propagators
of field $B^a_k$. (We take into account that considered diagrams
have only external lines corresponding to the field $A^a_\m$.) As a result, we obtain inequation
 \disn{37}{
\ga_m-\ga_a\le 2\be.
\nom}

We enumerate with the index $i$ the different subdiagrams (including the full diagram) of this diagram.
We denote as  $\ga_m^i,\ga_a^i,\be^i$ the corresponding characteristics of subdiagrams
(we suppose that external lines of a subdiagrams are not included into subdiagrams).
An inequality, similar to (\ref{37}), is not true for arbitrary subdiagrams,
because arbitrary subdiagrams can contain external lines corresponding
to the field $B^a_k$. However, this inequality is true for characteristics
of a part of the diagram not included in subdiagrams
(where external lines of subdiagrams are included to this part):
 \disn{38}{
(\ga_m-\ga_m^i)-(\ga_a-\ga_a^i)\le 2(\be-\be^i).
\nom}

The inequality $\om^i,\om_\p^i\le -1$ is a true for all diagrams of the class.
Hence, taking into account (\ref{29}),(\ref{29.1}), we can make a conclusion that inequality
 \disn{39}{
\om',\om'_\p\le \ga_a^i-\ga_m^i-1
\nom}
is true.
The indices of divergence for the estimating integral in formula (\ref{36})
are larger than corresponding indices for initial diagram (\ref{28}) by $2\be$,
because $\be$ copies of the propagators estimated by formula (\ref{35}) do not contribute to the indices.
This fact is true for subdiagrams as well. Hence, the estimated integral, in its turn, can be estimated (with accuracy up to logarithmic corrections) by
 \disn{40}{
\int\! dk\;\hat F(k)\le \ls\frac{1}{a}\rs^\al,\qquad
\al=\max\ls 0,\max_i\ls\ga_a^i-\ga_m^i-1+2\be^i\rs\rs.
\nom}
Here, we take the maximum among all subdiagrams (including the full diagram) and use the formula (\ref{39}).
Substituting estimation (\ref{40}) in (\ref{36}), we obtain estimation for the value of diagram in form
 \disn{41}{
\max\biggl((ma)^{\ga_m-2\be}a^{2\be-\ga_m+\ga_a},\;
\max_i\ls(ma)^{\ga_m-2\be}a^{1+2(\be-\be^i)-(\ga_m-\ga_m^i)+(\ga_a-\ga_a^i)}\rs\biggr).
\nom}
When taking into account (\ref{38}), this estimation can be changed as
 \disn{42}{
(ma)^{\ga_m-2\be}\,\max\ls a^{2\be-\ga_m+\ga_a},a\rs.
\nom}
Previously, we assumed that parameter $m$ increases as $1/a$ with accuracy up to logarithmic corrections.
Using this assumption and inequality (\ref{37}), we can obtain that either value of diagram tends to zero, either
 \disn{43}{
2\be-\ga_m+\ga_a=0,
\nom}
i.e. it behaves as $(ma)^{\ga_a}$. On this step of reasoning we assume that diagram
containing at least one propagator of field $B^a_k$ (hence $\be>0$) or at least one extra vertex (hence $\ga_a>0$).
But if $\be>0$ then (taking into account (\ref{43})) $\ga_a>0$
(we note that  $\ga_a$ is not equal to zero at $\ga_m=0$).
Therefore, if (\ref{43}) is true then $\ga_a>0$ is true too.
Therefore, value of the diagram is equal to zero in the limit $a\to0$
if we suppose that
product $(ma)$ logarithmically tends to zero.
It is possible if we assume
 \disn{44}{
m(a)=\frac{1}{a\ln\frac{1}{a\m}},
\nom}
where $\m$ is parameter of dimension of mass.

So, as a result, if we have to assume (\ref{44}) than all diagrams of this class (the diagrams with external lines type $A^a_\m$ and with $\om,\om_\p<0$ for full diagrams and all subdiagrams of these diagrams) either vanish in the limit $a\to0$ either tend to the value of corresponding diagrams of continuous gauge theory on group $U(N)$.

We can find the general form of such diagrams. We can analyze the contribution of propagators (\ref{25.1})-(\ref{25.3})
to the value $\om$,
which defined by formula (\ref{29}),
and contribution of vertices from expression (\ref{26}),(\ref{27})
and extra vertices from (\ref{22}) to one.
We have, that the contribution of all propagators of internal lines
(the propagator $\De_\m^{(A\La)ab}$ can be only external)
is equal to "$-2$", the contribution of vertices with three tails is equal to "$+1$"
and the contribution of vertices with four tails is equal to "$0$".
Hence, via standard reasoning, we can make conclusion that usual formula of continuous theory
 \disn{45}{
\om=4-L,
\nom}
is true for 1PI diagrams of considered theory.
Here, $L$ is number of external lines of the diagram. Also we can show that estimation
 \disn{46}{
\om_\p\le\om+L_--2n=4+L_--L-2n,
\nom}
is true for quantity $\om_\p$.
Here $n$ is number of loops of diagram and $L_-$ is number of external lines of the diagram
corresponding to the field $A_-$.
Note, that the 1PI diagrams with these external lines do not give contribution to Green functions,
but these diagrams give the contribution in vertex part $\Gamma$.
The analogue of Word identities is useful to be build in terms of these functions.
It is necessary to use such identities in construction of renormalization procedure for
considered theory.
Using
formulas (\ref{45}),(\ref{46}) we can see that the considered class of diagrams contains
diagrams with external lines type $A^a_\m$ and with $L>4$ and $L_-\le 2n$
(it
should be true for full diagram and for all subdiagrams of the diagram).
The obtained
in this section result is true for these diagrams.

The obtained result is also true for diagrams with divergence,
but only after applying some procedure of
subtraction, if in a result of this procedure the indices $\om,\om_\p$ become negative.
This fact is very useful for renormalization of the theory, because it is the main tool at the analysis of divergent parts of diagrams.

\section{The scheme of renormalization procedure}
In order to find correct form of the theory on transverse lattice which transforms
into usual gauge theory (which contains both abelian and nonabelian parts) in the limit $a\to0$,
we should find correct renormalization.
For this we should the counterterms to action (\ref{10}) which provide
the cancellation of the divergence parts of all diagrams. In this article, we only describe
the scheme of such renormalization procedure. Let's remind, that
we study only gauge invariant
"third" form of the theory without longitudinal divergences (see in Sect.~4).

The renormalization of nonabelian gauge theory (without introduction of lattice)
in light-like gauge $A_-=0$ using dimensional regularization was perfomed in the work Ref.~\cite{bas1} (see also Ref.~\cite{bas2}).
In such approach there is an additional problem for the analysis of the diagrams divergences
because Lorentz invariance is broken. Therefore, the divergence at the transverse momentum $p_1,p_2$ can be worse than the total divergence of diagram (see, for example, form of propagator (\ref{25.1})). As a result, divergent parts of diagrams can contain non-polynomial parts with respect to momenta.

As a first step of renormalization on transverse lattice, we should develop the subtraction procedure
for divergent parts of the diagrams. Using this procedure, we can obtain the dependence of the divergent parts of the diagrams from the external momenta of these diagrams. In perturbation theory the
lattice regularization lead to appearance of field $B_k$, modification of propagator of field $A_\m$
(which give usual expression for propagator
in limit $a\to0$), additional numbers of vertices and cutoff of transverse momentum, see Sect.~3.
Using the results of Sect.~5, we can expect that the divergent parts of diagrams are polynomials of second order or less on external momenta. We note that the order of these polynomials is determined by
usual dimensional reasons. The exception
can arise only for small number of diagrams with divergence along the transverse momentum.

Further, we can formulate an analog of Word
identities because theories are gauge invariant
(we note again, that the longitudinal divergences was removed by gauge invariant means, see in Sect.~4).
If we write the analogue of Word identities in terms of vertex part $\Gamma[A_\m,B_k]$ (which is a generation functional for 1PI diagrams) then these identities should contain 1PI diagrams with external lines corresponding to field $A_-$.
For this reason we introduce the field $\La$ in the theory (see (\ref{23})), instead of assuming $A_-=0$
(as in work Ref.~\cite{tmf99}), although these diagrams do not contribute to full Green functions of field $A_\m$.

Also, we should use remained space symmetries. They are rotational symmetry
in longitudinal space $k_3,k_4$ (for Euclidean form of theory), and the symmetry of discrete group of rotations
on $\pi/2$
in transverse space $k_1,k_2$, see equations (\ref{12})-(\ref{12.2}) and the text neer.
It should be noted that the divergent part can contain constant vectors $n_\m,\bar n_\m$ which are contained in the propagator (\ref{25.1}), because gauge was fixed. However,
this is possible only
if corresponding expression is invariant of the product of $n_\m$ by any constant, and, simultaneously, division of $\bar n_\m$ by the same constant (the propagator (\ref{25.1}) have this invariance).

As long as an analog of Word identities are satisfied, we can expect gauge invariance of counterterms,
which cancel the divergences of all diagrams.
These counterterms
should not break also the existent space symmetry. As the preliminary analysis show, this fact leads
to the usual stretch of field $A_C$, $A_k$, $B_k$ at the process of renormalization (we note that abelian and nonabelian parts of field stretch independently) and renormalization of coupling constant $g$. Also, certain renormalization factors
appear in different part items $L_3$ (see (\ref{12})) and $L_m$ of the action (\ref{10}).
It is possible that some terms that satisfy all symmetries and contain vectors $n_\m,\bar n_\m$ can appear in the action. This fact was found in Ref.~\cite{bas1}. That is why if we renormalize the theory, a number of unknown factors may appear in the action because total Lorentz invariance was
broken by transverse lattice and non-physical fields appeared.

Construction of the exact form of counterterms requires carefully performing the renormalization of theory on transverse lattice in according with the scheme described in this section. Such an investigation
is beyond of present paper, and it will performed in the next work.

\vskip 0.5em
{\bf Acknowledgments.}
The work was supported
by the Russian Ministry of Education, Grant No.~RNP.2.1.1/1575.
Authors express gratitude to V.A.~Franke and E.V.~Prokh\-va\-ti\-lov for useful discussion.

\end{document}